%% file: lcws21-2hdms.tex
\documentclass[a4paper,12pt]{article}
\pdfoutput=1
\usepackage{geometry}
\usepackage{amsmath,amssymb,amsfonts}
\usepackage{xcolor,graphicx,cite,soul}
\usepackage{array,booktabs,longtable}
\usepackage{caption,microtype}
\usepackage{subcaption}
\usepackage{hyperref}
\usepackage{datetime}
\usepackage{appendix}

\input{paperdef}

\graphicspath{{figs/}}
\allowdisplaybreaks
\begin{document}
\thispagestyle{empty}

\def\thefootnote{\fnsymbol{footnote}}

\begin{flushright}
IFT--UAM/CSIC-21-061\\
DESY 21-077
\end{flushright}

\mbox{}
\vspace{0.5cm}

\begin{center}

{\large\sc 
  {\bf A 96 GeV Higgs Boson in the 2HDMS:\\[.3em]
    \boldmath{$e^+e^-$} collider prospects}}%
\footnote{Talks presented at the International Workshop on Future
  Linear Colliders (LCWS2021),\\
  \mbox{}\hspace{5mm} 15-18 March 2021. C21-03-15.1} 


\vspace{1cm}

{\sc
S.~Heinemeyer$^{1,2,3}$%
\footnote{email: Sven.Heinemeyer@cern.ch}%
, C.~Li$^{4}$%
\footnote{email: cheng.li@desy.de}$^{**}$%
, F.~Lika$^5$%
\footnote{email: florian.lika@physnet.uni-hamburg.de}%
, G.~Moortgat-Pick$^{4,5}$%
\footnote{email: gudrid.moortgat-pick@desy.de}%
~and S.~Paasch$^4$%
\footnote{email: steven.paasch@desy.de}%
\footnote{Speakers}
}

\vspace*{.7cm}

{\sl
$^1$Instituto de F\'isica Te\'orica (UAM/CSIC), 
Universidad Aut\'onoma de Madrid, \\ 
Cantoblanco, 28049, Madrid, Spain

\vspace*{0.1cm}

$^2$Campus of International Excellence UAM+CSIC, 
Cantoblanco, 28049, Madrid, Spain 

\vspace*{0.1cm}

$^3$Instituto de F\'isica de Cantabria (CSIC-UC), 
39005, Santander, Spain

\vspace{0.1cm}

$^4$DESY, Notkestra\ss e 85, 22607 Hamburg, Germany

\vspace{0.1cm}

$^5$II. Institut f\"ur Theoretische Physik, Universit\"at Hamburg,\\
Luruper Chaussee 149, 22761 Hamburg, Germany

}

\end{center}

\vspace*{0.1cm}

\begin{abstract}
\noindent
The CMS collaboration reported a $\sim 3 \, \sigma$ (local) excess
at $96\;$GeV in the search for light Higgs-boson decaying into two photons.
This mass coincides with a $\sim 2 \, \sigma$ (local) excess in the
$b\bar b$ final state at LEP. We show an interpretation of these
possible signals as the lightest Higgs boson in the 2 Higgs Doublet Model
with an additional complex Higgs singlet (2HDMS). The interpretation is
in agreement with all experimental and theoretical constraints.
We concentrate on the 2HDMS type~II, which resembles the Higgs and
Yukawa structure of the Next-to Minimal Supersymmetric Standard Model.
We discuss the experimental prospects for constraining our explanation
at future 
$e^+e^-$ colliders, with concrete analyses based on the ILC prospects.
\end{abstract}


\def\thefootnote{\arabic{footnote}}
\setcounter{page}{0}
\setcounter{footnote}{0}

\newpage


\section{Introduction}
\label{sec:intro}

The Higgs boson discovered in 2012 by ATLAS and
CMS~\cite{Aad:2012tfa,Chatrchyan:2012xdj} is so far consistent with
the predictions of a Standard-Model~(SM) Higgs
boson~\cite{Khachatryan:2016vau} with a mass of $\sim 125 \gev$.
However, the experimental and theoretical uncertainties of the Higgs-boson
couplings are at the level of $\sim 10-20\%$,
leaving ample room for Beyond Standard-Model (BSM)
interpretations. Many theoretically well motivated extensions of the
SM contain an extended Higgs-boson sector. The additional Higgs boson
can have masses above, but also below $125 \gev$.

Searches for Higgs bosons below $125 \gev$ have been performed at LEP, the
Tevatron and the LHC. 
Two excesses have been seen at LEP and the LHC at roughly the same
mass, suggesting a possible common origin of both excesses via a new scalar.
LEP observed a $2.3\, \sigma$ local excess 
in the~$e^+e^-\to Z(H\to b\bar{b})$
searches\,\cite{Barate:2003sz}, at a mass scale of 
$\sim 98 \gev$. However, the mass
resolution is rather imprecise due to the
hadronic final state. The signal strength, 
i.e.\ the measured cross section normalized to the SM expectation
assuming a SM Higgs-boson mass at the same mass, 
was extracted to be $\mu_{\rm LEP}= 0.117 \pm 0.057$.

CMS searched for light Higgs bosons in the di-photon
final state. The results from Run II\,\cite{Sirunyan:2018aui} 
show a local excess of $\sim 3\, \sigma$ at
$\sim 96 \gev$, and a similar excess of $2\, \sigma$
at about the same mass~\cite{CMS:2015ocq} in the Run~I data.
Assuming gluon fusion to be the dominant production mode the
excess corresponds to $\mu_{\rm CMS}=0.6 \pm 0.2$.
Results from Run\,II from ATLAS
with~$80$\,fb$^{-1}$ in the di-photon final state turned out to be
still weaker than the corresponding CMS results (see, e.g., Fig.~1
in~\cite{Heinemeyer:2018wzl}).
Possibilities of how to
simultaneously explain both excesses by a common origin are discussed in the literature.

In \citere{Biekotter:2019mib} (see also
\citeres{Biekotter:2019drs,Biekotter:2020ahz}), it was analyzed that the
Two Higgs doublet model (2HDM) extended by a real singlet
(N2HDM)~\cite{Chen:2013jvg,Muhlleitner:2016mzt} can 
perfectly describe both excesses, while being in agreement with all
current theoretical and experimental constraints. It was shown
that the N2HDM type~II is best suited to explain the excesses. 
This points to further investigations into two directions. The Higgs sector
of supersymmetric models are naturally of type~II. Consequently, it is
of interest to test how well the more rigid structure of SUSY Higgs
sectors containting two doublets and (at least) one singlet can describe
the excesses. It was shown that indeed such models can accomodat the LEP
and the CMS excesses at the $1-1.5\,\sig$ level, while being in
agreement with all experimental 
data~\cite{Biekotter:2017xmf,Domingo:2018uim,Hollik:2018yek,Choi:2019yrv,Biekotter:2019gtq,Cao:2019ofo}
(for reviews see
\citeres{Heinemeyer:2018jcd,Heinemeyer:2018wzl,Richard:2020jfd}).
On the other hand, the singlets contained in the SUSY Higgs sector contain,
in contrast to the N2HDM, a complex singlet. Furthermore, an additional
$Z_2$ symmetry implemented into the
N2HDM~\cite{Chen:2013jvg,Muhlleitner:2016mzt} is not present in SUSY
models, which conversely obey an additional $Z_3$
symmetry. Consequently, it is of interest to analyze how well the 2HDM
with an additional complex singlet and the $Z_3$ symmetry imposed
(2HDMS) can describe the CMS and the LEP excess, while being in
agreement with all current theoretical and experimental constraints.

Here we present a first step into this direction. We sample a relevant
part of the 2HDMS parameter space, where it is found that the model can
perfectly accomodate both excesses~\cite{HLLMP}. We focus on
measurements of a future $e^+e^-$ collider of the $125 \gev$ Higgs-boson
couplings, and how they can reveal deviations from the SM. While such
measurements can be performed at any planned future $e^+e^-$ collider
(ILC, CLIC, FCC-ee or CEPC), concretely we use analyses for the
anticipated ILC precisions. A corresponding analysis for the N2HDM can
be found in \citere{Biekotter:2020ahz} (see \citere{Biekotter:2020cjs}
for the prospects at future LHC runs).


\section{The 2HDMS}
\label{sec:2hdms}

The Two-Higgs-Doublet Model with complex singlet (2HDMS) is an extension of
Two-Higgs-Doublets Model (2HDM) by an additional complex singlet field.  
The field content of the Higgs sector can be described as follows.
$\Phi_1$ and $\Phi_2$ are the $SU(2)$ doublets, and $S$ is the complex
gauge singlet. 
Imposing the $\mathbb{Z}_2$ Symmetry for two doublets can suppress the tree-level flavour-changing neutral currents (FCNC):
\begin{equation}
	\begin{split}
		\Phi_1\to\Phi_1,\qquad\Phi_2\to-\Phi_2,\qquad S\to S
	\end{split}
\end{equation}
Furthermore, one can assume that the Higgs structure of 2HDMS is NMSSM-like, corresponding to an additional $\mathbb{Z}_3$ Symmetry being
imposed on the Higgs potential. For this $\mathbb{Z}_3$ Symmetry, the
Higgs potential stays invariant under the transformation: 
\begin{equation}
	\begin{pmatrix}
		\Phi_1\\
		\Phi_2\\
		S
	\end{pmatrix}\to\begin{pmatrix}
		1&	&	\\
		&	e^{i2\pi/3}&	\\
		&	&	e^{-i2\pi/3}
	\end{pmatrix}\begin{pmatrix}
		\Phi_1\\
		\Phi_2\\
		S
	\end{pmatrix}
\end{equation}
The scalar potential is then given by
\begin{equation}
\begin{split}
  V=&m_{11}^2(\Phi_1^\dagger\Phi_1)+m_{22}^2(\Phi_2^\dagger\Phi_2)
  +\frac{\lambda_1}{2}(\Phi_1^\dagger\Phi_1)^2
  +\frac{\lambda_2}{2}(\Phi_2^\dagger\Phi_2)^2
  +\lambda_3(\Phi_1^\dagger\Phi_1)(\Phi_2^\dagger\Phi_2)\\   
&+\lambda_4(\Phi_1^\dagger\Phi_2)(\Phi_2^\dagger\Phi_1)+m_S^2( S^\dagger S)
  +\lambda'_1( S^\dagger S)(\Phi_1^\dagger\Phi_1)
  +\lambda'_2( S^\dagger S)(\Phi_2^\dagger\Phi_2)\\ 
  &+\frac{\lambda''_3}{4}( S^\dagger S)^2
  +\Big(-m_{12}^2\Phi_1^\dagger\Phi_2+\frac{\mu_{S1}}{6} S^3
  +\mu_{12} S\Phi_1^\dagger\Phi_2+\text{h.c.}\Big)
\end{split}
\label{eq:2hdmspot}
\end{equation}

After electroweak symmetry breaking, the scalar component of $\Phi_1$,
$\Phi_2$ and $S$ acquire the non trivial vaccum expectation values
(vevs). Thus the fields can be expanded around vevs and have the
following expressions: 
\begin{gather}
	\Phi_1=\begin{pmatrix}
		\chi_1^+\\\phi_1
	\end{pmatrix}=\begin{pmatrix}
		\chi_1^+\\v_1+\displaystyle{\frac{\rho_1+i\eta_1}{\sqrt{2}}}
	\end{pmatrix},\quad
	\Phi_2=\begin{pmatrix}
		\chi_2^+\\\phi_2
	\end{pmatrix}=\begin{pmatrix}
		\chi_2^+\\v_2+\displaystyle{\frac{\rho_2+i\eta_2}{\sqrt{2}}}
	\end{pmatrix},\notag\\
	S=v_S+\frac{\rho_S+i\eta_S}{\sqrt{2}}
\end{gather}
One can define $\tan\beta={v_2}/{v_1}$ as in the 2HDM.
The ``SM vev'' is then given by $v = \sqrt{v_1^2 + v_2^2} \approx 174 \gev$.

The charged Higgs sector of 2HDMS has the identical structure as the
charged Higgs sector of 2HDM. However, the additional singlet enter the
neutral Higgs sector and mix with two doublets for both CP-even sector
and CP-odd sector. Therefore, we have 3 scalar Higgs bosons $h_1$,
$h_2$, $h_3$, 2 pseudo-scalar Higgs bosons $a_1$, $a_2$, and the charged
Higgs boson $H^{\pm}$. We use the conventions of
$m_{h_1}<m_{h_2}<m_{h_3}$ and $m_{a_1}<m_{a_2}$. 

The CP-even sector is diagonalized as
\begin{equation}
\begin{pmatrix}
	h_1\\h_2\\h_3
\end{pmatrix}=R\begin{pmatrix}
	\rho_1\\ \rho_2\\ \rho_S
\end{pmatrix},\quad \mathrm{diag}\{m^2_{h_1},m^2_{h_2},m^2_{h_3}\}=R^T {M}^2_{S} R,
\end{equation}
where the symmetric $M^2_S$ is given by
\begin{eqnarray}
	M^2_{S11} &=& 2 \lambda _1 v^2 \cos^2 \beta + (m_{12}^2-\mu_{12}v_S)\tan\beta, \nonumber \\
	M^2_{S12} &=& (\lambda_3 +\lambda_4) v^2  \sin 2\beta - (m_{12}^2-\mu_{12}v_S), \nonumber \\ 
         M^2_{S13} &=& (2 \lambda^{'}_1 v_S \cos\beta+\mu_{12}\sin\beta)v, \nonumber  \\
	M^2_{S22} &=& 2 \lambda_2 v^2 \sin^2\beta+(m^2_{12} -\mu_{12}v_S)\cot \beta, \nonumber \\ 
	M^2_{S23} &=&  (2 \lambda^{'}_2 v_S \sin\beta  +\mu_{12}\cos\beta)v, \nonumber \\
	M^2_{S33} &=& \frac{\mu_{S1}}{2} v_S+\lambda^{''}_3 v_S^2-\mu_{12} \frac{v^2}{2 v_S} \sin2\beta
	\label{eq:msq}
\end{eqnarray}
with
\begin{equation}
	R=\begin{pmatrix}
		c_{\alpha_1}c_{\alpha_2}& s_{\alpha_1}c_{\alpha_2}& s_{\alpha_2}\\
		-s_{\alpha_1}c_{\alpha_3}-c_{\alpha_1}s_{\alpha_2}s_{\alpha_3}& c_{\alpha_1}c_{\alpha_3}-s_{\alpha_1}s_{\alpha_2}s_{\alpha_3}& c_{\alpha_2}s_{\alpha_3}\\
		s_{\alpha_1}s_{\alpha_3}-c_{\alpha_1}s_{\alpha_2}c_{\alpha_3}& -s_{\alpha_1}s_{\alpha_2}c_{\alpha_3}-c_{\alpha_1}{{s_{a}}}_3& c_{\alpha_2}c_{\alpha_3}
	\end{pmatrix},
	\label{eq:rot}
\end{equation}
where $\alpha_1$, $\alpha_2$ and $\alpha_3$ are the three mixing angles. 
Similarly, in the CP-odd sector the mixing angle $\al_4$ enters in addition. 
The set of input parameters is:
\begin{equation}
	\tan\beta,\quad\alpha_{1,2,3,4},\quad m_{h_1},\quad m_{h_2},\quad m_{h_3},\quad m_{a_1},\quad m_{a_2},\quad m_{H^\pm},\quad v_S
	\label{eq:inpmass}
\end{equation}
In our analysis we interpret the experimental excess at $\sim 96 \gev$
as the possible lightest scalar Higgs boson $h_1$, and we identify the
second lightest scalar Higgs $h_2$ as the SM-like Higgs at $\sim 125 \gev$.

The couplings of the Higgs bosons to the SM particles can be described
as follows. 
\begin{align}
  c_{h_i ff} &= \frac{g_{h_i ff}}{g_{H_\text{SM}ff}}~, \\
  c_{h_i VV} &= \frac{g_{h_i VV}}{g_{H_{\SM} VV}}~.
\end{align}
We focus here on the 2HDM type~II (see \refse{sec:intro}). Thus the
coupling modifiers for fermions are given by
\begin{align}
  c_{h_i tt} &= R_{i2}/\SB~, \non \\
\label{eq:fermioncoup}
  c_{h_i bb} &= R_{i1}/\CB~, \\
  c_{h_i \tau\tau} &= R_{i1}/\CB~. \non
\end{align}
Similarly, the coupling modifier for massive gauge bosons is given by
\begin{equation}
	c_{h_iVV}=c_{h_iZZ}=c_{h_iWW}=\cos\beta R_{i1}+\sin\beta R_{i2}~.
	\label{eq:bosoncoup}
\end{equation}


\section{Analysis details}

In the following subsection we list the relevant details for our
analysis: theoretical and experimental constraints, the $\chi^2$
contribution from the two excesses as well as the scanned 2HDMS
parameter space. 

\subsection{The constraints}
\label{sec:constraints}

In this section we give a brief description of the theoretical and
experimental constraints applied in our analysis to the 2HDMS.

\begin{itemize}

\item \textbf{Tree-Level perturbative unitarity:}\\
Tree-Level perturbative unitarity conditions ensures perturbativity up
to very high scales. This can be achieved by demanding the amplitudes of
the scalar quartic interactions, which are given by the eigenvalues of
the 2 $\rightarrow$ 2 scattering matrix, to be below a value of
$8\pi$. The calculation was carried out with a \Code{Mathematica}
package implemented in \Code{ScannerS}~\cite{scanners} and by following
the procedure of~\cite{Ho_ej__2006}.  

\item \textbf{Boundedness from below:}\\
Boundedness from below conditions ensures that the potential remains
positive when the field values approach infinity. 
The conditions can be found in~\cite{klimenko} and were adapted for the
2HDMS.

\item \textbf{Vacuum stability:}\\
An obvious condition is to require the EW vacuum to be the global
minimum (\textit{true vacuum}) of the scalar potential. In this case the
EW is absolutely stable. If the EW vacuum is a local minimum
(\textit{false vacuum}) the corresponding parameter region can still be
allowed if it is metastable. This is the case if the predicted life-time
is longer than the current age of the universe. Any configuration with a
life-time shorter than the age of the universe is considered unstable.  
For our study we used \Code{Evade}~\cite{Hollik_2019,Ferreira_2019,evade}.

\item \textbf{Higgs-boson rate measurements:}\\
We use the public code
\Code{HiggsSignals v.2.5.1}~\cite{Bechtle:2013xfa,Bechtle:2014ewa,Bechtle:2020uwn}
to verify that all generated points agree with currently
available measurements of the SM Higgs-boson. \Code{HiggsSignals} uses a
statistical $\chi^2$ analysis of the SM-like Higgs-boson of a given
model and compares it to the measurements of the Higgs-boson signal
rates and masses at Tevatron and LHC.

\item \textbf{BSM Higgs-boson searches:}\\
The public code
\Code{HiggsBounds v.5.9.0}~\cite{Bechtle:2008jh,Bechtle:2011sb,Bechtle:2013wla,Bechtle:2015pma,Bechtle:2020pkv}
provides $95\%$ confidence level exclusion limits of all important
direct searches for charged Higgs bosons and additional neutral
Higgs-bosons.  

\item \textbf{Flavor physics observables:}\\
The charged Higgs sector of the 2HDMS is unaltered with respect to the
general 2HDM, and we can take over the corresponding 
constraints from flavor physics. 
The most important bounds, see \citere{Arbey:2017gmh}, come from
BR$(B_s\to X_s \gamma)$, constraints on $\Delta M_{B_s}$ from
neutral B-meson mixing and BR$(B_s\to\mu^+\mu^-)$.
These give a lower limit for the charged Higgs mass
$m_{H^\pm}\geq 650 \gev \gev$~\cite{Haller:2018nnx}.

\item \textbf{Electroweak precision observables:}\\
Constraints from electroweak precision observables (EWPO) can be
expressed in terms of the parameters $S$, $T$ and
$U$~\cite{Peskin:1990zt,Peskin:1991sw}. If BSM physics enters  
mainly through gauge boson self-energies, as it is the case for the
extended Higgs sector of the 2HDMS, these parameters can give a good
approximation to capture the constraints from EWPO. In the 2HDMS the
parameter $T$ is the most sensitive and has a strong correlation to
$U$. Contributions to U can therefore be
dropped~\cite{Funk:2011ad,Biekotter:2019kde}. For our scan we require
the prediction of the $S$ and $T$ 
parameters to be within the 95\% confidence level interval, corresponding to
$\chi^2 = 5.99$ for two degrees of freedom. The calculations were
carried out by \texttt{SPheno-4.0.4}~\cite{Porod_2012,Porod_2003}.  

\end{itemize}


\subsection{The experimental excesses}
\label{sec:excesses}

The experimental excesses at both LEP and CMS could be translated to the
following signal strengths: 
\begin{align}
  \mu^\text{exp}_\text{LEP} &=
  \frac{\sigma(e^+ e^-\to Z \phi\to Z b \bar{b})}
       {\sigma(e^+ e^-\to Z H^0_\text{SM}\to Z b \bar{b})} = 0.117 \pm 0.05~,\\
   \mu^\text{exp}_\text{CMS} &=
   \frac{\sigma(p p\to \phi\to \gamma\gamma)}
        {\sigma(p p\to H^0_\text{SM}\to \gamma\gamma)} = 0.6 \pm 0.2~,
\end{align}
where the $H^0_\text{SM}$ is the SM Higgs-boson with the rescaled mass
at the same range as the new scalar particle $\phi$ at $\sim 96 \gev$.

We interpreted the new scalar $\phi$ as the lightest CP-even Higgs-boson
$h_1$ of the 2HDMS. Its signal strengths can be calculated by the
following expressions in the narrow width approximation: 
\begin{align}
\label{eq:mulep}
  \mu^\text{theo}_\text{LEP} &=
  \frac{\sigma_\text{2HDMS}(e^+ e^-\to Z h_1)}
       {\sigma_\text{SM}(e^+ e^-\to Z H^0_\text{SM})}
       \times\frac{\text{BR}_\text{2HDMS}(h_1\to b\bar{b})}
                  {\text{BR}_\text{SM}(H^0_\text{SM}\to b\bar{b})}
      = |c_{h_1 VV}|^2 \frac{\text{BR}_\text{2HDMS}(h_1\to b\bar{b})}
                           {\text{BR}_\text{SM}(H^0_\text{SM}\to b\bar{b})} \\
  \mu^\text{theo}_\text{CMS} &=
  \frac{\sigma_\text{2HDMS}(g g\to h_1)}
       {\sigma_\text{SM}(g g\to H^0_\text{SM})}
       \times \frac{\text{BR}_\text{2HDMS}(h_1\to \gamma\gamma)}
              {\text{BR}_\text{SM}(H^0_\text{SM}\to \gamma\gamma)}
      = |c_{h_1 tt}|^2 \frac{\text{BR}_\text{2HDMS}(h_1\to \gamma\gamma)}
                          {\text{BR}_\text{SM}(H^0_\text{SM}\to \gamma\gamma)}
	\label{eq:mucms}
\end{align}
The effective couplings of $c_{h_1 VV}$ and $c_{h_1 tt}$ in
\refeqs{eq:bosoncoup} and (\ref{eq:fermioncoup}). The corresponding BRs are
evaluated with \texttt{SPheno-4.0.4}~\cite{Porod_2012}~\cite{Porod_2003}. 

The $\chi^2$ contribution from these two excesses can then be calculated as
\begin{equation}
  \chi_\text{CMS-LEP}^2 =
  \left(\frac{\mu_\mathrm{LEP}^\text{theo}-0.117}{0.057}\right)^2
  +\left(\frac{\mu_\mathrm{CMS}^\text{theo}-0.6}{0.2}\right)^2
\label{eq:chi2excess}
\end{equation}
The SM, in which no scalar at $\sim 96 \gev$ is present thus receives a
$\chi^2$ penalty of $\sim 13.2$. In our analysis we will only consider
points in the 2HDMS which are in agreement with the two excesses at the
$1\,\sig$ level, i.e.\ with $\chi_\text{CMS-LEP}^2 \le 2.3$.


\subsection{Parameter scan}
\label{sec:scan}

Following the N2HDM interpretation~\cite{Biekotter:2019kde}, we focus on
the type-II Yukawa structure and the low $\tan\beta$ region. 

As discussed before, in
our analysis we interpret the experimental excess at $\sim 96 \gev$
as the possible lightest scalar Higgs boson $h_1$, and we identify the
second lightest scalar Higgs $h_2$ as the SM-like Higgs at $\sim 125 \gev$.
We furthermore made the pedagogical guess that the lightest CP-odd
Higgs $a_1$ is singlet dominant, which corresponds to the condition of
$\sin^2\alpha_4>\frac{1}{2}$. To avoid decay $h_2 \to a_1 a_1$ we
explore an $m_{a_1}$ range of $200 \gev$ to $500 \gev$.
Following the preferred ranges in the N2HDM~\cite{Biekotter:2019kde}, we
scan the following parameter space:
\begin{align}
	&m_{h_1}\in \{80,\;93\}\;\text{GeV}, &m_{h_2}&\in \{110,\;126\}\;\text{GeV},&m_{h_3}&\in \{650,\;1000\}\;\text{GeV},\notag\\
    &m_{a_1}\in \{200,\;500\}\;\text{GeV}, &m_{a_2}&\in \{650,\;1000\}\;\text{GeV},&m_{H^\pm}&\in \{650,\;1000\}\;\text{GeV},\notag\\
	&\tan\beta\in \{1,\;6\},&\alpha_4&\in \{1.25,\;\frac{\pi}{2}\},&v_S&\in \{100,\;2000\}\,\text{GeV}\notag\\
	&\frac{\tan\beta}{\tan\alpha_1}\in \{0, 1\}, &\alpha_2&\in \pm\{0.95,\;1.3\}, &|\sin(\beta&-\alpha_1-|\alpha_3|)|\in \{0.98, 1\}
\end{align}


\section{Results: prospects for \boldmath{$e^+e^-$} colliders}
\label{sec:results}

The results of our parameter scans in the type II of the 2HDMS~\cite{HLLMP},
indicate that this model configuration
can accommodate both excesses simultaneously, while being in
agreement with all considered constraints described above.
This was to be expected because of the similarity of the 2HDMS and the
N2HDM, where the latter was shown to perfectly fit the excesses
previously~\cite{Biekotter:2019mib,Biekotter:2019drs,Biekotter:2020ahz,Biekotter:2020cjs}. 

The particle $h_1$ is dominantly singlet-like, and acquires its
coupling to the SM particles via the mixing with the SM Higgs boson
$h_2$. Thus, the presented scenario will be experimentally accessible
in two complementary ways. Firstly, the new particle $h_1$ can be
produced directly in collider experiments (see, e.g.,
\citere{Drechsel:2018mgd}), an avenue we will not
pursue here. Secondly, deviations of the
couplings of $125\, \gev$ Higgs boson $h_2$ from the SM predictions are
present. 
Currently, the uncertainties of the measurement of the coupling
strengths of the SM-like Higgs boson at the LHC are still
large~\cite{Khachatryan:2016vau,ATLAS-CONF-2018-031,Sirunyan:2018koj}.
In the future, even tighter constraints are expected from the HL-LHC with
$3000$\,fb$^{-1}$ integrated luminosity~\cite{Dawson:2013bba}. 
Finally, a future linear $e^+ e^-$ collider, for instance the ILC, could
improve the precision measurements of the Higgs boson couplings
with unprecedented precision (see, e.g.. \citere{Bambade:2019fyw}).%
\footnote{Similar
results can be obtained for CLIC, FCC-ee and CEPC. We will focus on the ILC 
prospects here following \citere{Biekotter:2019mib}.}%
~We compare our scan points to the expected precisions of
the LHC and the ILC as they are reported
in Refs.~\cite{Bambade:2019fyw,Cepeda:2019klc},
neglecting possible correlations of the coupling modifiers.

\begin{figure}
  \centering
  \includegraphics[width=0.60\textwidth]{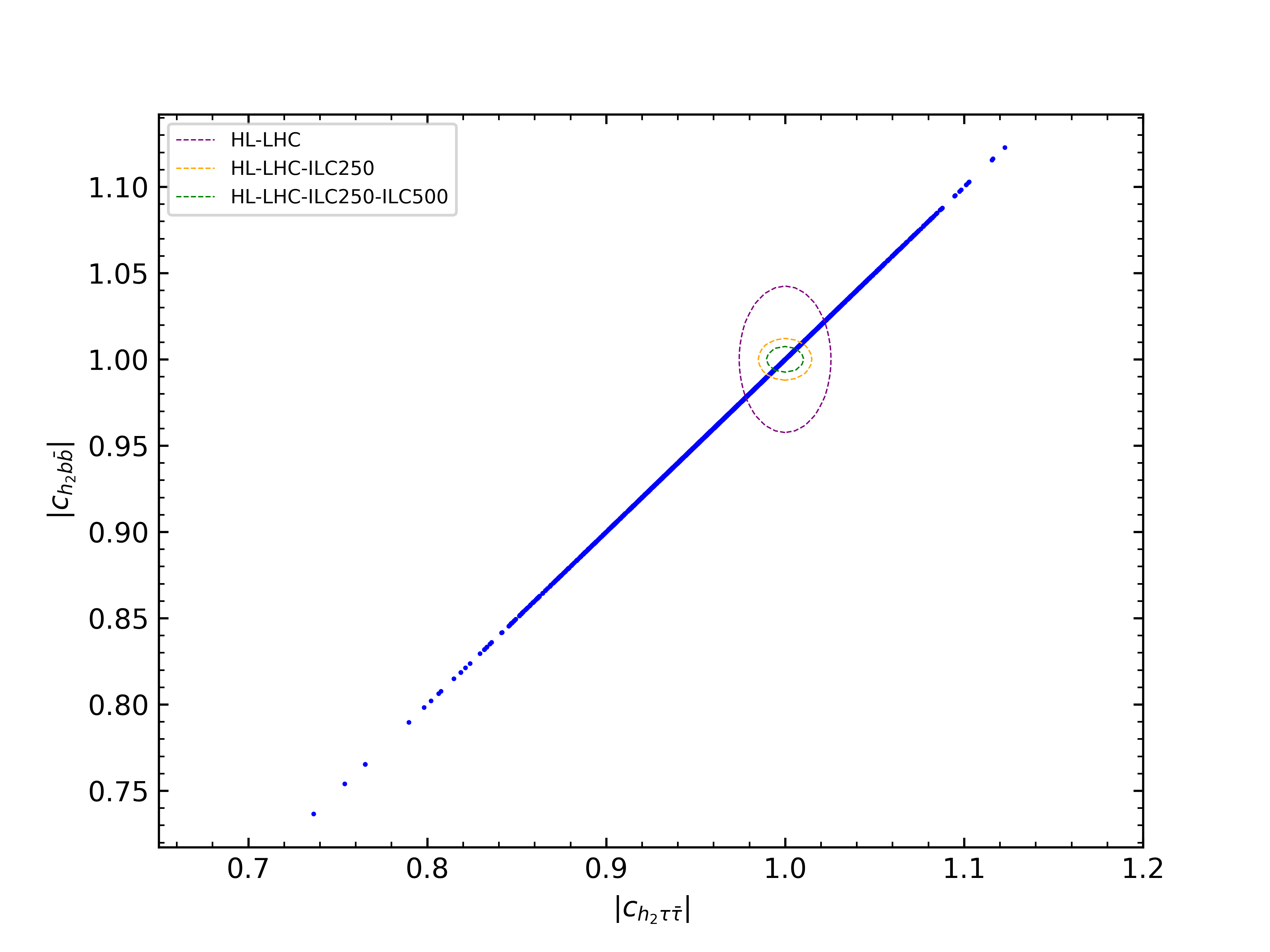}\\
  \includegraphics[width=0.60\textwidth]{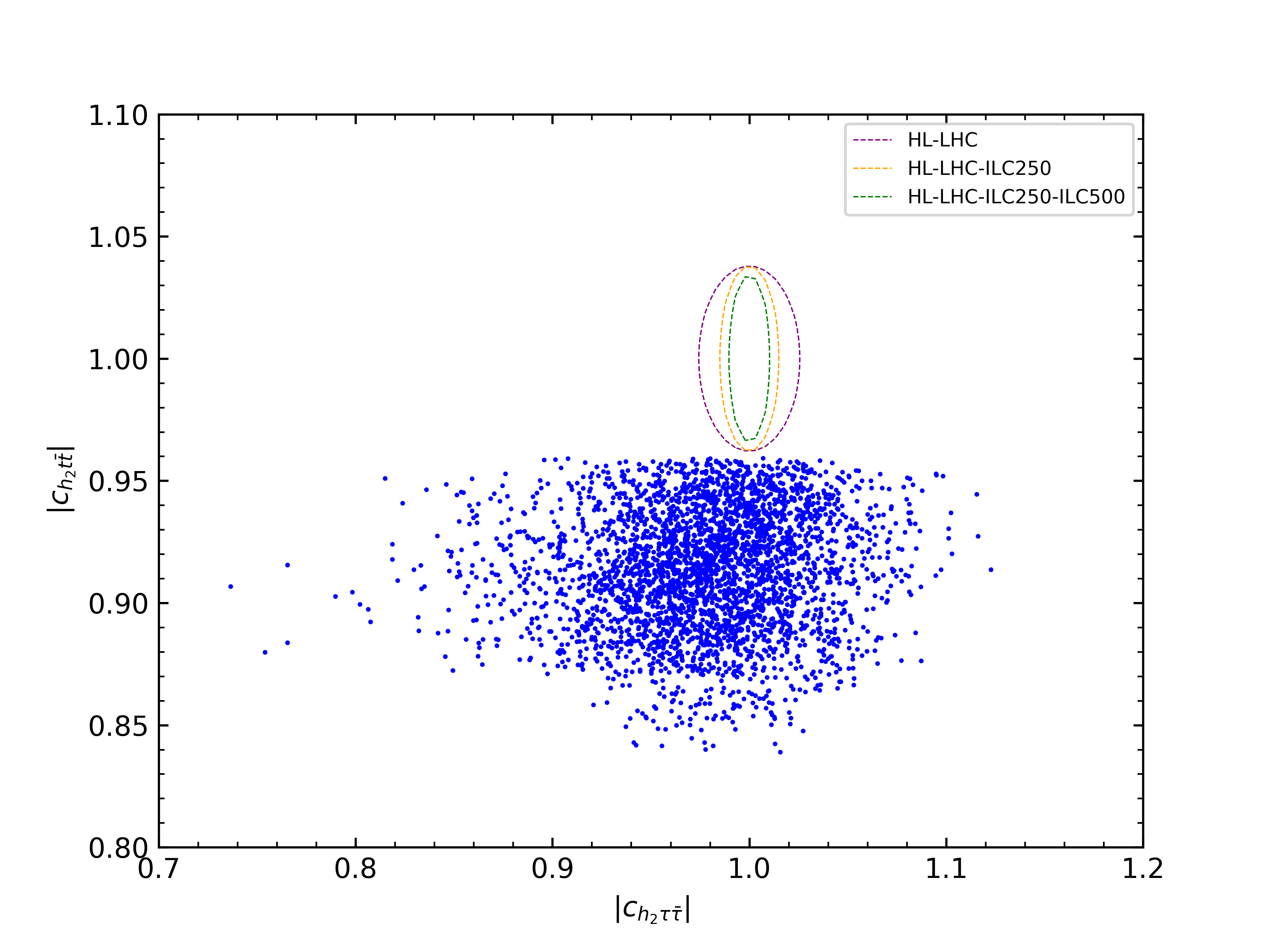}\\
  \includegraphics[width=0.60\textwidth]{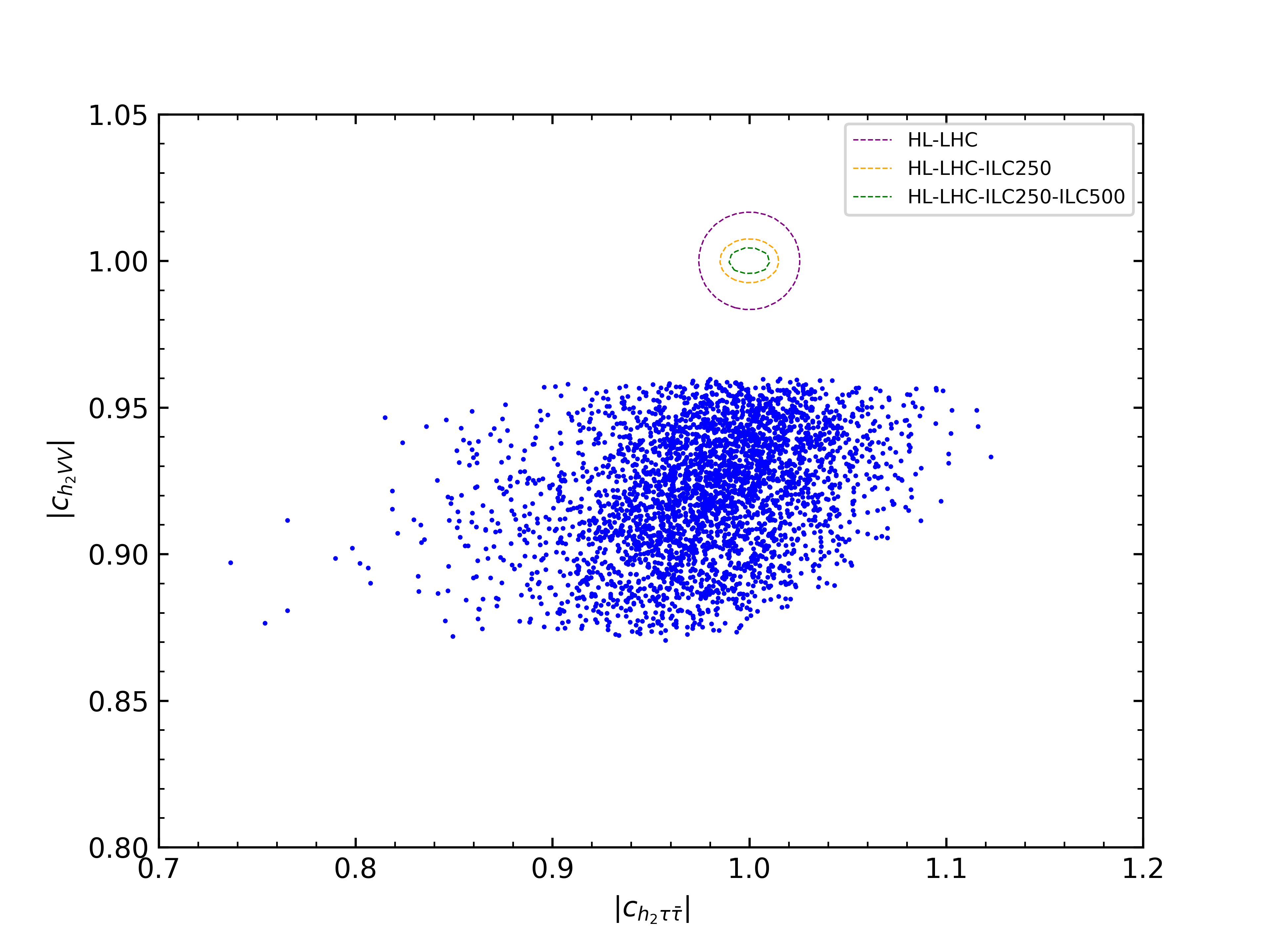}
  \caption{Prospects for the Higgs coupling measurements at the HL-LHC
  and the ILC (see text). The upper, middle and lower plot show the
  planes of $|c_{h_2\tau\tau}|-|c_{h_2t\bar t}|$,
  $|c_{h_2\tau\tau}|-|c_{h_2b\bar b}|$,$|c_{h_2\tau\tau}|-|c_{h_2VV}|$}
  \label{fig:h2-coup}
\end{figure}

In Fig.~\ref{fig:h2-coup} we plot the coupling modifier
of the SM-like Higgs boson $h_2$ to $\tau$-leptons, $c_{h_2\tau\tau}$ 
on the horizontal axis against the coupling coefficient to
$b$-quarks, $c_{h_2b\bar b}$ (top), to $t$-quarks, $c_{h_2t\bar t}$
(middle) and to the massive SM gauge bosons, $c_{h_2VV}$ (bottom).
These points passed all the experimental and theoretical constraints
and have $\chi^2_\text{CMS-LEP} \le 2.3$. We include several future precisions
for the coupling measurements. It should be noted that they are centered
around the SM predictions to show the potential to discriminate
the SM from the 2HDMS. The magenta ellipse in each plot shows the
expected precision of the measurement of the coupling coefficients at
the $1 \, \sigma$-level at the HL-LHC from Ref.~\cite{Cepeda:2019klc}.
The current uncertainties and the HL-LHC analysis are based on the
coupling modifier, or $\kappa$-framework. These modifiers are then
constrained using a global fit to projected HL-LHC data assuming no
deviation from the SM prediction will be found. 
We use the uncertainties given under the assumptions that
no decay of the SM-like Higgs boson to BSM particles is present,
and that current systematic uncertainties will be reduced in addition
to the reduction of statistical uncertainties due to the increased statistics.
The green and the orange ellipses show the corresponding expected
uncertainties when the HL-LHC results are combined with projected
data from the ILC after the $250\gev$ phase and
the $500\gev$ phase, respectively, taken from Ref.~\cite{Bambade:2019fyw}.
Their analysis is based on a pure effective field theory calculation,
supplemented by further assumptions to facilitate the combination with
the HL-LHC projections in the $\kappa$-framework. In particular, in
the effective field theory approach the vector boson couplings can
be modified beyond a simple rescaling. This possibility was excluded
by recasting the fit setting two parameters related to the couplings
to the $Z$-boson and the $W$-boson to zero
(details can be found in \citere{Bambade:2019fyw}).

The future precision in the Higgs coupling measurement to quarks can be
summarized as follows. The uncertainty of the coupling to $b$-quarks
will shrink below $4\%$ at the HL-LHC and below $1\%$ at the ILC.
The coupling to $\tau$-leptons an uncertainty at the level of $2\%$
is expected at the HL-LHC. Here, the ILC could reduce this uncertainty 
to below $1\%$. The coupling to $t$-quarks are
expected to reduce the uncertainty by roughly a factor of three at the
HL-LHC, employing, however, theoretical assumptions.
At the ILC the accuracy of these couplings cam be substantially improved to
be model-independent.

In the top plot the 2HDMS points lie on a diagonal line, because in
type~II models the coupling to leptons and to down-type quarks scale identically.
This diagonal goes through the SM prediction
($c_{h_2\tau\tau} = c_{h_2bb} = 1$).
While current constraints on the SM-like Higgs-boson
properties allow for large deviations of the couplings of up
to $40\%$, the parameter space of our scans that is still allowed
will be significantly reduced by the expected measurements at the HL-LHC
and the ILC.\footnote{Here one should 
keep in mind the theory input required in the (HL-)LHC analysis.}
The couplings of the $125 \gev$ Higgs boson to $t$~quarks, shown in the
middle plot of \reffi{fig:h2-coup}, on the other
hand show a deviation from the SM prediction for all points. The
deviation ranges from $\sim 1.5\,\sig$ to about $\sim 4\,\sig$. 

Finally, in the lower plot of Fig.~\ref{fig:h2-coup}, where we show the
absolute value of the coupling modifier of the SM-like Higgs boson
w.r.t.\ the vector boson couplings $|c_{h_2 V V}|$, 
the parameter points of the 2HDMS show a deviation from the SM
prediction substantially larger than the projected experimental uncertainty
at HL-LHC and ILC.
A $2-3\,\sig$ deviation from the SM prediction
is expected 
with HL-LHC accuracy. At the ILC a deviation of more than $5\,\sigma$
would be visible. 
Consequently, the 2HDMS explaining the two excesses in the low-mass
Higgs boson searches at LEP and CMS at $\sim 96 \gev$,
can be distinguished from the SM by the measurements of the $125 \gev$
Higgs-boson couplings at the ILC.


\section{Conclusions}
\label{sec:conclusions}

The possibility of the 2HDMS explaining both the CMS and the LEP excesses
at $\sim 96 \gev$ 
simultaneously offers interesting prospects to be probed experimentally.
Here we have focused on the possibilities to measure the properties of
the $125 \gev$ Higgs boson at future $e^+e^-$ colliders. Concretely, we
used the projected accuracies of the ILC running at $\sqrt{s} = 250 \gev$
and $500 \gev$, similar results are expected for CLIC, FCC-ee or
CEPC. 
We analyzed such points that are in agreement with all experimental and
theoretical constraints and have $\chi^2_\text{CMS-LEP} \le 2.3$
(i.e.\ they accomodate the two excesses at the $1\,\sig$ level).
These points show a deviation from the prediction of the SM Higgs boson
that can clearly be detected by the ILC coupling measurements. While
part of the 2HDMS spectrum reproduces the SM coupling to $b$~quarks, the
coupling to $t$~quarks deviates between $\sim 1.5$ and $\sim 4\,\sig$
from the SM prediction. By measuring the coupling to the massive SM
gauge bosons, a deviation of more than $5\,\sig$ is expected.
Thus, the 2HDMS explaining the excesses in the Higgs boson searches at
LEP and CMS at $\sim 96 \gev$, can clearly be distinguished from the SM
at the ILC.


\subsection*{Acknowledgements}

We thank T.~Biek\"otter for very constructive discussions on the N2HDM results.
The work of S.H.\ is supported in part by the
MEINCOP Spain under contract PID2019-110058GB-C21 and in part by
the AEI through the grant IFT Centro de Excelencia Severo Ochoa
SEV-2016-0597.
C. L., G.M.-P.\ and S.P.\  acknowledge the support by the Deutsche
Forschungsgemeinschaft (DFG, German Research Foundation) under Germany’s
Excellence Strategy – EXC 2121 ``Quantum Universe – 390833306.'' 


\end{document}

%% file: paperdef.tex
\newcommand{\non}{\nonumber}

\newcommand\Code[1]{\ensuremath{\texttt{#1}}}

\newcommand\al{\alpha}

\newcommand\SB{s_\beta}
\newcommand\CB{c_\beta}

\newcommand\ReDiag{\mathop{%
  \raise .5pt\hbox{[}%
  \widetilde{\mathrm{Re}}%
  \raise .5pt\hbox{]}}}
\newcommand\ReOffDiag{\mathop{%
  \raise .5pt\hbox{$\llbracket$}%
  \widetilde{\mathrm{Re}}%
  \raise .5pt\hbox{$\rrbracket$}}}

\newcommand\SM{\mathrm{SM}}

\newcommand\refeqs[1]{Eqs.~(\ref{#1})}

\newcommand\refse[1]{Sect.~\ref{#1}}

\newcommand\citere[1]{Ref.~\cite{#1}}
\newcommand\citeres[1]{Refs.~\cite{#1}}

\newcommand{\gev}{\,\, \mathrm{GeV}}

\newcommand{\sig}{\sigma}

\def\reffi#1{\mbox{Fig.~\ref{#1}}}

\definecolor{Orange}{named}{orange}
\definecolor{Purple}{named}{purple}
\definecolor{Lightblue}{cmyk}{0.9,0.1,0.1,0.3}
\definecolor{dgelborange}{cmyk}{0.,0.3,0.5, 0.}
\definecolor{Lila}{rgb}{0.5,0.,1}